\documentclass[twocolumn,amsmath,amssymb,floatfix]{revtex4}

\usepackage{bm}
\usepackage{amsmath}
\usepackage{epsf}
\newcommand{\beq}{\begin{equation}}

\newcommand{\eeq}{\end{equation}}
\newcommand{\beqa}{\begin{eqnarray}}
\newcommand{\eeqa}{\end{eqnarray}}

\newcommand{\etal}{{\it et al. }}

\def\simlt{\lesssim}
\def\simgt{\gtrsim}

\newcommand{\ApJL}{Astrophys. J Lett.}
\newcommand{\ApJ}{Astrophys. J}

\newcommand{\PRD}{Phys. Rev. D}
\newcommand{\MNRAS}{Mon. Not. R. Astron. Soc.}
\newcommand{\ARAA}{Ann. Rev. Astron. Astrophys.}

\newcommand{\aut}[2]{{#2.\ #1,}}
\newcommand{\saut}[2]{{#2.\ #1,}}
\newcommand{\laut}[2]{{#2.\ #1,}}
\newcommand{\refs}[6]{#2, {\bf #3},  {#4} (#5).}

\newcommand{\mybib}[2]{\bibitem{#2}}
\begin{document}

%@twocolumnfalse\endcsname
\title{Redshifting Rings of Power}
\author{Wayne Hu$^{1}$ \& Zoltan Haiman$^{2}$}
\affiliation{
{$^{1}$}Center for Cosmological Physics, Department of Astronomy and Astrophysics, 
and Enrico Fermi Institute, University of Chicago, Chicago IL 60637\\
{$^{2}$}Department of Astronomy, Columbia University, 550 West 120th Street, New York NY 10027
}

\begin{abstract}
\baselineskip 11pt
The cosmic microwave background (CMB) has provided a precise 
template for features in the linear power spectrum: the matter-radiation turnover,
sound horizon drop, and acoustic oscillations. 
In a two dimensional power spectrum in redshift and angular
space, the features appear as distorted rings, and yield simultaneous, purely
geometric, measures of the Hubble parameter $H(z)$ and angular
diameter distance $D_A(z)$ via an absolute version of the Alcock-Paczynski test.  
Employing a simple Fisher matrix tool, we
explore how future surveys can exploit these rings of power for dark
energy studies.  High-$z$ CMB determinations of $H$ and $D_A$ are best
complemented at moderate to low redshift ($z \simlt 0.5$) with a population of
objects that are at least as abundant as clusters of galaxies.  We
find that a sample similar to that of the ongoing SDSS Luminous Red Galaxy (LRG) survey 
can achieve
statistical errors at the $\sim 5\%$ level for $D_A(z)$ and
$H(z)$ in several redshift bins.  This, in
turn, implies errors of $\sigma(w)=0.03-0.05$ for a constant dark
energy equation of state in a flat universe.  Deep galaxy cluster surveys 
such as the planned South Pole Telescope (SPT) survey, can extend this
test out to $z \sim 1$ or as far as redshift followup is available. 
We find that the expected constraints are at the
$\sigma(w)=0.04-0.08$ level, comparable to galaxies and complementary in
redshift coverage.
\end{abstract}
\maketitle

\section{Introduction}

Cosmic microwave background (CMB) measurements have now established the sound
horizon and horizon at matter-radiation equality 
as standard rulers for cosmology.  Current errors on the absolute 
scales are approximately $2\%$ and $8\%$ respectively \cite{Speetal03} 
and will continue to improve as higher angular resolution data further resolve
the morphology of the acoustic peaks.  
From the angular scale subtended by these rulers, the CMB provides
a comparably precise measurement of the angular diameter distance $D_A$ 
to the epoch of recombination.  

These rulers appear in the matter power spectrum as a smooth turnover 
at the matter-radiation horizon, a sharper drop at the sound horizon, and
a series of acoustic oscillations at harmonics 
of the sound horizon \cite{HuSug96,EisHu98}.
The acoustic features are close analogues to the CMB peaks and 
cause variations of order $10\%$ in power.  They
are preserved out to nearly 
the non-linear scale at any given redshift \cite{MeiWhiPea99}. 
The whole set of
features does not evolve with redshift in the linear regime in the absence
of a significant massive neutrino component to the dark matter.
Matching of the observed features with the template provided
by the CMB provides cosmological distance measures $D_A(z)$ based on the
same physics and method as the CMB 
\cite{EisHuTeg99a,CooHuHutJof01}.  
There is, in fact, more information in a three dimensional redshift
survey since the redshift dimension measures the evolution
of the Hubble parameter $H(z)$ and so the dark energy density evolution directly \cite{Eis03}.  

These features appear
as {\it rings} in the two dimensional angular and
redshift space.  At $z=0$, the angular diameter
distance $D_A$ and the Hubble parameter $H$ depend only on the current
expansion rate.  Here the rings are circular and the measurement of their
location returns the Hubble constant $H_0$ \cite{EisHuTeg99a}.
Between $0 < z \simlt 1$, the scalings of $D_A$ and $H$ depend on the dark energy
evolution whereas for $z \simgt 1$ they are expected to return to 
the matter-dominated scalings provided by the CMB.
The relative distortion between
the redshift and angular dimensions provides the Alcock-Paczynski 
\cite{AlcPac79} test for dark energy.   Here, we show that the absolute distortion can
be measured from the absolute calibration of the standard rulers.

Recently there have been several studies of the utility of acoustic
features to constrain cosmology at $z=1-3$ emphasizing the
extended range of the linear regime and the large volumes
encompassed by surveys with moderate 
angular dimensions
(e.g. a few hundred square degrees) 
\cite{Eis03,BlaGla03,Lin03}.  High-$z$ surveys would provide important consistency checks
with the high-$z$ CMB and the Hubble parameter determinations as
well as limit any residual dark energy component at those epochs.  
However, given the CMB determination, the most important regime to probe
is at moderate to low redshift, since this provides the largest lever--arm
in distance.  

The quantitative tools developed
to address these issues 
have focused on the angle-averaged $P(k)$ which entangles
and degrades 
information on $D_A$ and $H$ \cite{BlaGla03}.  
In contrast, the separately-developed general tools for analyzing 
the information in
redshift surveys at cosmological distances are complete
but computationally costly to implement
\cite{MatSza02}.  
This difficulty has prevented a full exploration of
the use of two-dimensional power spectrum rings 
given a CMB power spectrum template and the requirements that 
their measurement places on surveys. 
Here we generalize the approximate mode 
counting estimates for the power spectrum \cite{FelKaiPea94} 
to the two dimensional space
and quantify the information through a Fisher matrix approach.  

We begin in \S \ref{sec:rings} with a discussion of the cosmological
distortion of power spectrum rings.  We continue in \S \ref{sec:estimate} with
the mode counting technique for estimating the
capabilities of redshift surveys.  In \S \ref{sec:toy}, we describe two
mock surveys for illustration purposes, 
one based on the ongoing SDSS luminous red galaxy (LRG) 
survey at intermediate redshifts and the other
based on a deep galaxy cluster survey
such as the planned South Pole Telescope (SPT) survey \cite{CarHolRee02}. 
We examine potential
cosmological constraints on distances, the Hubble parameter and the
dark energy in \S \ref{sec:distance}. Their dependence on survey and other
assumptions is explored in  in \S \ref{sec:priors}.  We
discuss the results in \S  \ref{sec:discussion}.  

Throughout this paper, we take as the fiducial cosmology 
a flat $\Lambda$CDM model with baryon density $\Omega_b h^2=0.024$, matter 
density $\Omega_m h^2 =0.14$,
scalar slope $n_s=1$, dark energy density in 
units of the critical density $\Omega_{\rm DE}=0.72$ 
(or Hubble constant $H_0=100 h$ km s$^{-1}$ Mpc$^{-1}$ with $h=0.72$), 
 initial curvature $\delta_\zeta = 5.07 \times 10^{-5}$ at $k=0.05$ Mpc$^{-1}$
(or present normalization $\sigma_{8}=0.9$ and reionization optical depth $\tau=0.17$), 
and a constant dark energy equation of state $w=p_{\rm DE}/\rho_{\rm DE}=-1$.  These values are
consistent with recent determinations from WMAP \cite{Speetal03}.

\section{Rings of Power}
\label{sec:rings}

Geometrical distortions at cosmological distances are
described by the Friedmann-Robertson-Walker spatial line element
\begin{equation}
ds^2= a^2(dD^2 + D_A^2 d\Omega)\,.
\end{equation}
The metric elements are related to the observable redshift as
$a=(1+z)^{-1}$ for the scale factor,
$d D = dz/H$ for the radial distance $D$, 
and $D_A = R \sin (D/R)$ for the angular diameter distance,
where the radius of
curvature $R = H_0^{-1} \sqrt{\Omega_{\rm tot}-1}$
and $\Omega_{\rm tot}$ is the total density in units of the critical density.
All distances are in comoving coordinates.
The Hubble parameter is given by the Friedmann equation as
\begin{equation}
H^2(z) = {8\pi G \over 3} \rho_{\rm tot}(z) - {1 \over (a R)^2}\,,
\end{equation}
with $H_0 = H(z=0)$. Since the conversion from the observable angular 
and redshift space coordinates
to physical coordinates depends on the metric, a ``standard ruler'' of a known
physical size can be used to measure cosmology.

\begin{figure}[tb]
\centerline{\epsfxsize=3.00truein\epsffile{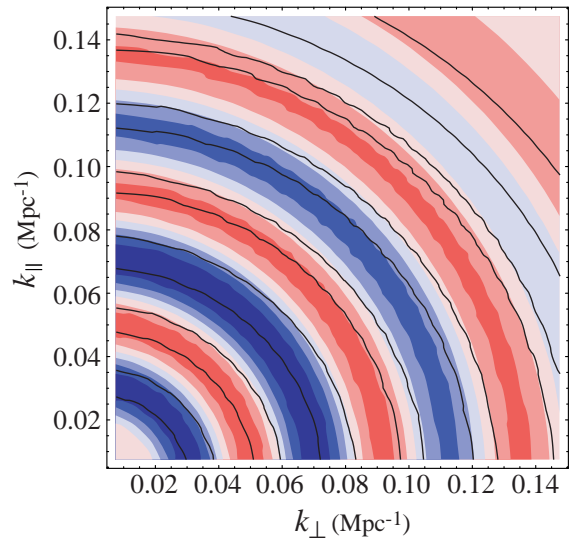}}
\caption{Acoustic rings in the two dimensional power spectrum 
$P_s/\bar P_s - 1$ 
with a smooth component $\bar P_s$ \cite{EisHu98} removed
to highlight the
features; shaded contours are spaced by $0.02$.  
The locations of the features 
are preserved in the presence of linear redshift space distortions here
at $z=0.45$ and $b=3.5$.  Cosmology distorts the rings here shown with
$w=-2/3$ and $\Omega_{\rm DE}=0.62$, $h=0.61$ (lines surround extrema of the oscillations)
which preserves the
CMB-determined high-$z$ $D_A$ and $H$.  Jaggedness reflects our $k$-cell discretization.}
\label{fig:rings}
\end{figure}

Suppose now that we have a survey of some biased tracer of the mass, 
effectively at some redshift $z_s$.  The two point correlation or 
power spectrum of the objects acts as the standard ruler.  
Since power spectrum modes are usually
quoted in units of inverse length scale, let us choose the fiducial cosmology
for the conversion
\begin{eqnarray}
k_\perp^{\rm fid} &=& {\ell \over D_A^{\rm fid}(z_s)}\,, \nonumber\\
k_\parallel^{\rm fid} &=& {2\pi \over \lambda_z} {\Delta z \over \Delta D} 
\approx {2\pi \over \lambda_z} H^{\rm fid}(z_s) 
\,,
\end{eqnarray}
where $\lambda_z$ is the radial wavelength in redshift and
$\ell$ is the angular wavenumber or multipole.  Note that in our 
fiducial flat cosmology $D_A=D$. 
The true modes being
probed by a given $\ell$ and $\lambda_z$ are 
\begin{eqnarray}
k_\perp &=& s_\perp  k_\perp^{\rm fid}\,,\nonumber\\
k_\parallel &=& s_\parallel  k_\parallel^{\rm fid}\,,
\end{eqnarray}
where the shift parameters are
\begin{eqnarray}
s_\perp &=& {D_A^{\rm fid} \over D_A}\,, \nonumber\\
s_\parallel &=& {{\Delta D}^{\rm fid} \over {\Delta D}} \approx { H \over H^{\rm fid} }\,.
\label{eqn:shiftparams}
\end{eqnarray}
In the linear regime, the power spectrum of the tracer objects 
reflect the underlying mass power spectrum $P(k)$ modified by 
redshift space distortions as \cite{Kai87}
\begin{eqnarray}
P_s(k_\perp,k_\parallel) &=& \left[1 + \beta \left( {k_\parallel \over k}\right)^2
\right]^2 b^2 P(k)\,, \nonumber\\
k^2  & =& k_\perp^2 + k_\parallel^2 \,,
\end{eqnarray}
where $b$ is the linear bias assumed to be scale independent 
(e.g. \cite{Col93}).   Deep in the linear regime,
the distortion parameter
\begin{equation}
\beta = {1 \over b} {d\ln D_{\rm grow} \over  d\ln a}\,,
\end{equation}
where $D_{\rm grow}$ is the linear growth
rate; we alternatively consider $\beta$ to be a free parameter 
to reflect uncertainties in the distortion approaching the non-linear regime.
With prior knowledge of the
underlying form of $P(k)$ from the CMB, the shifting in the
observational domain measures the
angular diameter distance $D_A$ and Hubble parameter $H$ through
Eqn.~(\ref{eqn:shiftparams}).

The redshift power spectrum is shown for the fiducial model in Fig.~\ref{fig:rings}.
Note that the locations of the rings remain undistorted
in the presence of the smooth linear 
redshift space distortion but not under a change in the
cosmology.  Note further that the usual Alcock-Paczynski test utilizes only 
the departure of the rings from perfect circles \cite{AlcPac79}, regardless of their radius.  
The information we utilize here is different: it is the departure
from the absolute size and shape of the rings expected from the CMB.
Hence $D_A$ and $H$ can in principle be measured independently. Moreover
in the complete absence of template information from the CMB, cosmological constraints
on the dark energy become substantially weaker and subject to uncertainties in the 
correction of redshift space distortions \cite{MatSza02}.

\section{Error Estimates}
\label{sec:estimate}

To estimate potential constraints on cosmology from the rings, we begin
by considering statistical errors on the measured two-dimensional power spectrum.
Statistical errors arise from the finite number of spatial samples of the clustering
and that of the tracer objects contributing shot noise. These will depend on the effective volume probed by the sample which we
will parameterize 
by the solid angle $A_s$,
redshift $z_s$ and redshift extent $\Delta z_s$
of the survey. 

The survey dimensions define a set of fundamental modes or equivalently a
fundamental unit in 3 dimensional $k$-space.  
Each of these units can be
considered an independent measurement of the power spectrum and so the
net error on a larger cell in $k$-space over which the power spectrum
is measured is simply $(2/N)^{1/2}$, where $N$ is the number of
fundamental units in the cell (see e.g. \cite{Tegetal98}).  The shot noise
from the finite number density of the tracers increases the fractional
errors by $(1+1/\bar n P_s)$ to give a net error of \cite{FelKaiPea94}
\begin{eqnarray}
\left( {\Delta P_s \over P_s} \right)^2 = {2 \over V_k V_{\rm eff}} \,,
\end{eqnarray}
where
\begin{equation}
V_{\rm eff}(k) = \int dV_s \left[ {\bar n(z_s) P_s(k) \over 1 + \bar n(z_s) P_s(k)} 
\right]^2
\label{eqn:FKPweight}
\end{equation}
is the effective volume probed and
\begin{equation}
V_k = {2 \pi \Delta\, (k_\perp^2) \Delta k_\parallel \over (2\pi)^3}
\end{equation}
is the $k$-space volume of the double ring of positive and negative
$k_\parallel$.  Note that the total $k$-volume we utilize
below is cylindrical and hence a factor of 1.5 larger than the 
volume of a corresponding
sphere with a radius $k_{\rm max}= k_{\perp, {\rm max}} =
k_{\parallel, {\rm max}}$ (effectively adopted in previous
one--dimensional studies).  For a roughly constant number density, the
variance scales as $1/V_s$ or $1/A_s$.  The covariance between bands
is negligible so long as $\Delta k_\perp \gg (D_A^{\rm fid}
A_s^{1/2})^{-1}$, $\Delta k_\parallel \gg (\Delta z_s /H^{\rm
fid})^{-1}$ \cite{Tegetal98}.

\begin{figure}[tb]
\centerline{\epsfxsize=3.00truein\epsffile{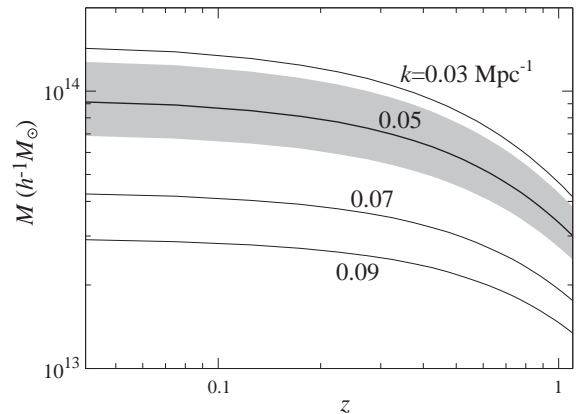}}
\caption{Tracers of power at the peak wavelengths of the
acoustic rings with a signal to shot noise $\bar n P_{s}= 2$ (range of $1.5-2.5$ shaded 
for the $k=0.05$ ring).   Tracers are
quoted in terms of the parent halo mass threshold which parameterizes
the number abundance and bias of the tracer population.}
\label{fig:abundance}
\end{figure}

Power spectra error estimates can be converted into error estimates on cosmological
parameters.
Given a set of parameters $p_\mu$ the Fisher matrix 
\begin{equation}
F_{\mu\nu}=  \sum_i 
{\partial \ln k_\perp^2 k_\parallel P_{s i} \over \partial p_\mu}
 { V_{k\,i} V_{{\rm eff}\,i} \over 2}
{\partial \ln k_\perp^2 k_\parallel P_{s i}
\over \partial p_\nu} 
\label{eqn:fisher}
\end{equation}
approximates the variance of the parameter estimates as $\sigma^2(p_\mu) = ({\bf F}^{-1})_{\mu\mu}$.  
Note that fractional
errors on $k_\perp^2 k_\parallel P(k)$ are the same as for $P(k)$ for sufficiently small cells and so this prescription follows from simple error 
propagation.

The subtleties in Eqn.~(\ref{eqn:fisher}) lie in the choice of
the observable $k_\perp^2 k_\parallel P_{s i}$.
Here $i$ indexes an array of cells in two dimensions that are
evaluated with fixed observable 
$k_\perp^{\rm fid}$, $k_\parallel^{\rm fid}$ and
hence $k_\perp$ and $k_\parallel$ that shift with cosmological parameters
$p_\mu$.  This prescription is the high-$z$ analogue of
bands in $h$ Mpc$^{-1}$.  Likewise the weighting $k_\perp^2 k_\parallel
\propto s_\perp^2 s_\parallel$ 
reflects the observable angular and redshift space 
variance and is the high-$z$ analogue of 
quoting power spectra in units of $h^{-3}$ Mpc$^3$.   Thus some sensitivity
in this test is coming from the change in the volume element with
cosmology. This information is degenerate with an 
amplitude change in the power spectrum at the given redshift.  

\section{Mock Surveys}
\label{sec:toy}

Tracer objects must have a sufficient number density 
and clustering strength for a measurement of the power spectrum in the presence
of shot and other noise sources.   
Integration time is best spent in tracing
a large volume with the lowest abundance objects that suffice.  
The power spectrum error estimates of the previous section provides
a simple criteria for the choice of astrophysical tracers such as
galaxies and galaxy clusters.

For definiteness we will mock up the tracer objects as dark matter halos
to a limiting mass in the theoretical mass function 
(\cite{Jenetal01}, eqn. B3)
that approximately matches the number densities expected
from a given observation.  This prescription also fixes the
bias through the halo prescription \cite{MoWhi96} as modified by \cite{SheTor99} for the improved
mass function.   Note that the larger bias of the high mass 
objects partially offsets their lower number density.
Shown in Figure \ref{fig:abundance} is the 
halo mass scale at which the signal-to-noise 
$\bar n P_s(k_\perp=k,k_\parallel=0) =2$ for the fiducial
cosmology for $k$-values corresponding to the extrema of the acoustic oscillations
\cite{BlaGla03}.  
For intermediate redshifts, objects with the number density and bias of low mass
clusters and groups suffice.   

Let us therefore consider
two mock surveys, one based on galaxies and the other on clusters.
We choose parameters that are roughly similar to the 
SDSS main and LRG sample \cite{Eisetal01}
and a deep cluster survey such as the planned SPT
survey \cite{CarHolRee02}. 
Specifically we take the number density of LRGs to be $\bar n=1\times 10^{-4}
h^3$Mpc$^{-3}$ at $z=0.3$ \cite{Eisetal01} and so mock the population with halos of
$M > 10^{13.5}$ $h^{-1} M_\odot$ and hence an average bias $b \approx 2$.
We divide the sample into three bins of $\Delta z=0.1$ from $z=0.1-0.4$.  
We take the main survey to consist of galaxies with 
$\bar n= 3 \times 10^{-3}$ $h^3$Mpc$^{-3}$ corresponding to
a population with $M > 10^{12.1}$ $h^{-1} M_\odot$ and $b \approx 1$ in 
a single volume from $z=0-0.1$.  We take both samples to cover
a sky area of $A_s=10000$ deg$^2$.  
Note that the LRG population is well-placed for measuring the
acoustic rings both in terms of number density and redshift.

We take the cluster survey to cover
$A_s=4000$ deg$^2$, to have a constant mass threshold of 
$M > 10^{14.2}$ $h^{-1} M_\odot$
\cite{MajMoh02}, and to extend from 
$0.1 < z < 1.3$ binned in steps of $\Delta z=0.2$.  
We will assume that the cluster survey has followup redshifts on all 
clusters ($N\approx 25000$). We explore the precision required in the redshifts and degradation 
suffered from lack of high-$z$ followup 
($N\approx 14000$; $0.1<z<0.7$) in \S \ref{sec:priors}.
For both mock surveys we approximate the number 
densities to be constant across the
redshift bin evaluated at the angular diameter distance midpoint.

To be conservative, we
marginalize the bias and hence the information contained in the growth
of structure.  This procedure isolates the geometric aspect of the 
test so that a misestimate in the tracer bias simply scales the signal-to-noise and does
not bias the results.  We leave a full consideration of the cosmological
information in the power spectrum to a separate work \cite{HaiHu03} 
(see also \cite{MajMoh03}).

We take $29^2$ linearly spaced $k$-cells from $k_{\perp,\parallel}
=0.005-0.15$ Mpc$^{-1}$ defining a cylinder in 3-dimensional $k$-space. 
The maximum $k$ in each direction hits the non-linear
scale at $z=0$ and the measurements will be somewhat degraded by 
non-linear corrections to the redshift space distortion 
and the washing out of features in the mass power spectrum \cite{MeiWhiPea99};
we explore the dependence of the results on $k_{\rm max}$ in \S \ref{sec:priors} below.

We account for uncertainties in the CMB template matter power spectrum
by allowing ($\Omega_b h^2$, $\Omega_m h^2$, $n_s$) to vary from the
fiducial model.   Current CMB uncertainties are at the level
$\sigma(\ln \Omega_b h^2)=0.04$, $\sigma(\ln \Omega_m h^2)=0.08$, $\sigma(n_s)=0.03$ 
\cite{Speetal03}.  
We take $(0.01, 0.01, 0.01)$
as a projection of future CMB constraints implemented by the addition of
$(F_{\rm prior} )_{\mu\nu}= \delta_{\mu\nu} \sigma(p_\mu)^{-2}$ to the Fisher matrix of
Eqn.~(\ref{eqn:fisher}).  
We discuss the effect of weakening these priors in \S \ref{sec:priors}.

For reference, the Planck satellite can in principle achieve errors of
$(0.009,0.0065,0.004)$ \cite{Hu01c}.   Moreover the
combination of WMAP with 
high resolution ground and balloon based measurements should also
be able to reach this level eventually. 
When implementing Planck-specific priors in the consideration of
dark energy constraints, we employ the full Fisher matrix forecast 
rather than the diagonal approximation.

Aside from these three parameters, we also include uncertainties
due to the initial normalization $\ln \delta_\zeta$, the reionization optical depth
$\tau$ and the tensor-scalar ratio $T/S$, when considering prior knowledge
from the CMB.  The remaining parameters in the Fisher matrix are then
$\ln \beta$, $\ln b$ and the dark energy parameters. We take  weak
priors $\sigma(\ln \beta)=\sigma(\ln b)=0.4$ unless
otherwise stated.

\begin{figure}[tb]
\centerline{\epsfxsize=3.00truein\epsffile{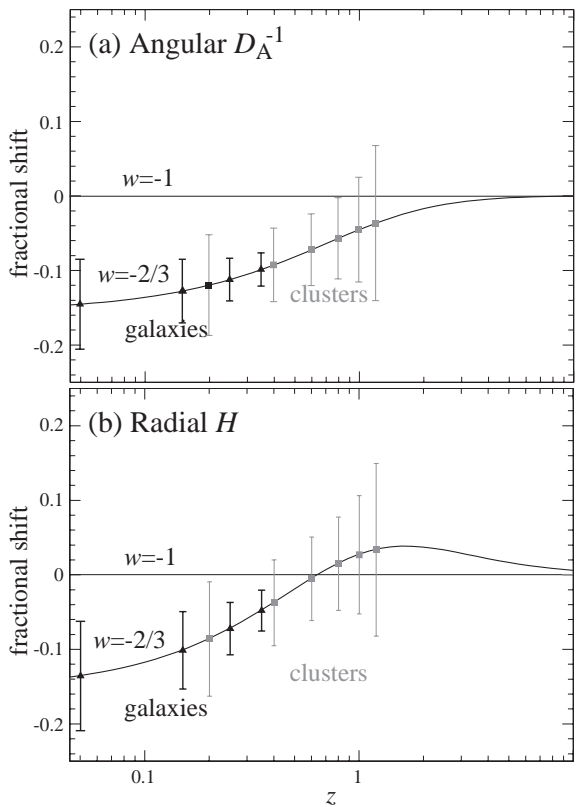}}
\caption{Projected constraints on the fractional shifts ($s_{\perp,
\parallel}-1$) in (a) $D_A$ and (b) $H$ from 
the fiducial model given the galaxy and cluster surveys specified in \S \ref{sec:toy}.  
Constraints are marginalized over the redshift space distortion 
$\beta$ and the bias parameter in each bin separately.
The high-$z$ tails of the $D_A$ and $H$ curves are constrained by the CMB, currently
to a precision of $\sim 2\%$ and $\sim 4\%$ ($\sigma(\ln\Omega_m h^2)=0.08$); 
projections assume a future improvement to $\sigma(\ln \Omega_m h^2)=
\sigma(\ln \Omega_b h^2)=\sigma(n_s)=0.01$ for the power spectrum template but the
difference is not substantial.}
\label{fig:shift}
\end{figure}

\section{Cosmological Constraints} 
\label{sec:distance}

A cosmology parameterized by a discrete set of shift parameters 
$s_\parallel(z_s)$ and $s_\perp(z_s)$ away from a fiducial cosmology
returns the distance and Hubble parameter as $D_A=D_A^{\rm fid}/s_\perp$ and
$H = H^{\rm fid} s_\parallel$ at the given redshift.  This parameterization is valid for
any model of dark energy evolution, including those involving spatial curvature 
$R\ne 0$, and is a complete description for 
geometrical tests.  The underlying assumption is that the
dark energy remains smooth on survey scales and hence does not affect the shape
of the power spectrum.

The projected constraints on the shift parameters for the galaxy and
cluster mock surveys
are shown in Fig.~\ref{fig:shift}. To eliminate the information coming from
redshift space distortions and the growth of structure, we
marginalize $\beta$ and the bias $b$ with weak priors. 

To understand the efficacy of the projected constraints for dark energy studies
recall that the fiducial cosmology is chosen so that $D_A^{\rm fid}(z_*)$ 
and $H^{\rm fid}(z_*)$ hit the central value of the WMAP results 
\cite{Speetal03}.  Thus $s_\parallel$, $s_\perp$ are constrained to lie 
near unity
at $z_*$ the recombination redshift.  In terms of cosmological
parameters, a fixed $H^{\rm fid}(z_*)$ implies a fixed $\Omega_m h^2$.

Deviations due to the influence of the dark energy mainly appear at low redshift.
In a simple constant-$w$ parameterization of the dark energy, the high
redshift constraint requires $\Omega_{\rm DE}$ to decrease as $w$ increases since
\begin{equation}
\rho_{\rm DE}(z) = {3 H_0^2 \over 8\pi G} \Omega_{\rm DE} (1+z)^{3(1+w)}\,.
\end{equation}
In Fig.~\ref{fig:shift}, we show an example satisfying this constraint
with $w=-2/3$ and hence $\Omega_{\rm DE}=0.62$.
This constraint tends
to make dark energy effects disappear more rapidly as $z$ is increased than
variations say at a fixed $\Omega_{\rm DE}$.
Note that at some redshifts the shift in the parallel and perpendicular
directions have the opposite sign.  An angle-averaging of the power
spectrum \cite{BlaGla03} would recover an average shift of
\begin{equation}
\bar s \approx  {2\over 3}s_\perp + {1\over 3} s_\parallel
\end{equation}
and hence this procedure both degrades the 
signal and complicates the interpretation for
a model of the dark energy with arbitrary evolution.

Since $s_\parallel(z=0)$ and $s_\perp(z=0)$ 
both return $H_0/H_0^{\rm fid}$, the low redshift side directly
measures the Hubble constant.  
In fact the best single redshift at which to complement the CMB is $z=0$ because
a change in $\Omega_{\rm DE}$ implies a change in $h$ in a flat universe
since $\Omega_m h^2 = (1-\Omega_{\rm DE}) h^2$ 
is fixed by the high-$z$ Hubble parameter determination.
Note the large deviation between the curves in Fig.~\ref{fig:shift}
at $z=0$ reflecting an $h=0.61$ and the rapid return to the fiducial
values for $z \simgt 1$.  

The galaxy sample therefore provides the best
complement to the CMB constraint at high-$z$. It 
can yield a highly significant separation between the $w=-2/3$ 
and $w=-1$ fiducial models and potentially even 
an internal detection of the evolution in $D_A^{-1}$ and $H$ due to the dark energy.
Even though the clusters do not have the number density to be optimal 
for this test, their ability to probe intermediate redshifts would be invaluable
for constraints on dark energy evolution.

\begin{figure}[tb]
\centerline{\epsfxsize=3.00truein\epsffile{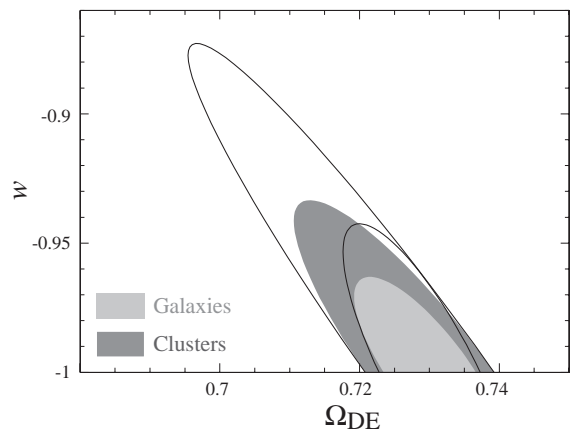}}
\caption{Projected $68\%$ CL constraints on $w$, $\Omega_\Lambda$ from the
galaxy and cluster surveys described in \S \ref{sec:toy} in combination
with Planck priors on the $P(k)$ template and $D_A(z_*)$, $H(z_*)$.  
Filled ellipses
represent constraints with a global bias amplitude marginalized and with
$\beta$ given by linear theory; open ellipse marginalize a scaling factor to $\beta$ and also the
bias separately at each redshift.}
\label{fig:ellipse}
\end{figure}

Joint constraints on the dark energy equation of state
$w$ depend on how well the CMB fixes the high-$z$ tail
of the curves in Fig.~\ref{fig:shift}.  In Fig.~\ref{fig:ellipse} we 
employ a projection of Planck capabilities from \cite{Hu01c}
and combine them with the constraints from the mock galaxy and cluster
surveys redshift slices by adding the Fisher matrices. 
Here we assume a dark energy parameterization given by $(w,\Omega_{\rm DE})$ 
in a flat universe.
Filled ellipses represent constraints with a single overall bias
parameter $b$ marginalized and with $\beta$ given by the linear theory
prediction.  We discuss sensitivity to these assumptions in \S 
\ref{sec:priors}.  
Note that the geometric constraint from the
CMB peaks alone follows a pure $D_A(z_*)$ degeneracy curve in 
the $(\Omega_{\rm DE},w)$ plane and
so the ability to measure $\Omega_{\rm DE}$ and $w$ comes from the
rotation of the degeneracy curves provided by measurements of $D_A$ and $H$
at lower redshifts.

For the galaxy survey, the 
net constraint on a constant $w$ is $\sigma(w)=0.024$ and
$\sigma(\Omega_{\rm DE})= 0.007$. 
For the cluster survey,
the net constraint is still a comparable 
$\sigma(w)=0.04$ and $\sigma(\Omega_{\rm DE}) = 0.013$ 
because of the higher bias and extended $\Delta z = 1.2$ assumed.
At the high-$z$
end, constraints become degenerate with those of the CMB itself for models close
to the fiducial model.  
Again it is important to
bear in mind that tracking the return
to the CMB values in $D_A$ and $H$ would be important for any future
detection of a dynamical dark energy component. 
This is especially true for models with a redshift dependent equation of
state $w(z)$ where the dark energy can contribute to the energy
density even at high-$z$.

\section{Exploring Assumptions}
\label{sec:priors}

The Fisher approach enables us to explore efficiently the physical origin of the
cosmological information and the dependence of projected
constraints on our assumptions.  The main assumptions are the
detectability of the acoustic rings, the extent of the linear regime, 
the determination of redshifts for the objects, the number density of tracers, and
the prior constraints.  

\begin{figure}[tb]
\centerline{\epsfxsize=3.00truein\epsffile{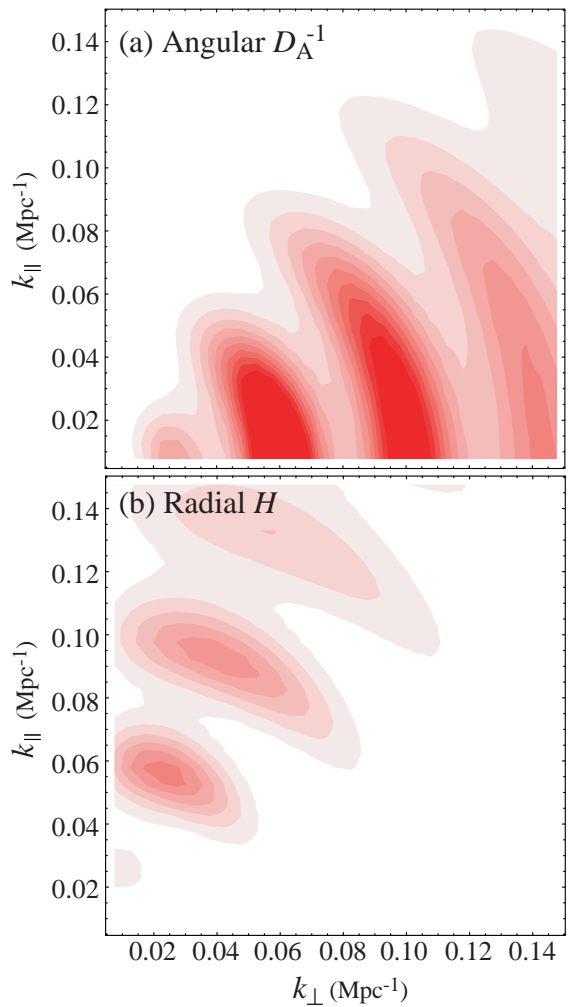}}
\caption{Contributions to the Fisher matrix showing the localization of
information on $D_A$ (or $s_\perp$) and $H$ (or $s_\parallel$) in the
$(k_\perp,k_\parallel)$ plane.  Here $z=0.4$ and we have taken clusters
with a mass threshold of $10^{14.2}$ $h^{-1} M_\odot$ ($\bar n=8 \times 10^{-6}$
$h^3$ Mpc$^{-3}$, $b = 3.4$).  
Notice the concentration of information
at the rise and fall to the acoustic rings, the drop in information at high
$k$ due to shot noise and the $k$-space volume weighting of the
sample variance which pulls the $s_\parallel$ information away from
the vertical axis.}
\label{fig:fisher}
\end{figure}

It is instructive to consider first the source of the cosmological information.
In Fig.~\ref{fig:fisher} we show the contribution to the Fisher matrix
for the parameters $s_\parallel$ and $s_\perp$.  
Here we take the mock cluster survey at $z=0.4$.
For illustrative purposes, we have first subtracted out the 
constant factor in the derivatives provided by the volume
weighting $k_\perp^2 k_\parallel \propto s_\perp^2 s_\parallel$ as this 
is degenerate with the 
normalization (see discussion following Eqn.~[\ref{eqn:fisher}]).  Note 
that this choice still only gives a rough quantification of the
independent information since any constant factor is degenerate with 
normalization.

Not surprisingly
$s_\perp$ (or $D_A$) gains most of its information from modes
nearly perpendicular to the line of sight 
and with $k\simlt 0.1$ due to the shot noise at high $k$.  The influence
of the acoustic rings is clearly visible with the largest effect being
between the extrema.
However even the information
in $s_\parallel$ is weighted toward non-zero $k_\perp$ due to increase
in sample variance from the small $k$-space around $k_\perp=0$.
Furthermore, estimates of $s_\perp$ and $s_\parallel$ come from nearly
non-overlapping $k$-cells and hence yield nearly independent determinations
of $D_A$ and $H$.

A quantification of the amount of information coming from the acoustic rings 
versus the broader features at the matter-radiation and sound horizons is
an important for the planning of surveys. The recovery of sharp features
places stringent requirements on the survey geometry and argues 
for a large contiguous survey region.  
To address this question
we replace the numerically calculated $P(k)$ from an Einstein-Boltzmann
code with a smooth fit 
which retains both the matter-radiation bend and sound horizon
drop but not the oscillations \cite{EisHu98}.  
Constraints on $H$ and $D_A$ then degrade by a factor of $1.7$
for the LRG galaxies.  

The acoustic rings become even more important if the $P(k)$ template
is not well determined by the CMB.
Weakening the prior constraints
on the template parameters $(\Omega_b h^2,\Omega_m h^2, n_s)$
to the current level (which may also be viewed as
a rough proxy for future uncertainties due to massive neutrinos, scale-dependent
bias and running of the tilt \cite{EisHuTeg99b,HaiHu03}) causes a degradation of a factor of
$\sim 1.1$ for the LRG galaxies
 in the presence of the acoustic rings
but a factor of 2.2 in their absence.  The acoustic rings therefore
help make this geometric test robust to uncertainties in 
the power spectrum template.

The extent to which the information in the acoustic rings
can be recovered with broad windows
from a non-contiguous region of sky is beyond the 
scope of the mode counting approximation used here.  It is best addressed
with the full pixel correlation 
matrix formalism \cite{MatSza02}.    Here we simply note
that with prior knowledge of the power spectrum template from the CMB, 
accurate knowledge of the window functions may in some circumstances suffice.

Because much of the information is coming from the acoustic rings,
the most critical prior assumption we have made is the extent of linear regime.
Changing $k_{\rm max}$ from $0.15$ to $0.075$ Mpc$^{-1}$ degrades
the distance and Hubble errors for the LRG 
galaxies by a factor of $\sim 2$.
These degradation factors are less pronounced for clusters
where the higher rings are lost in the shot noise (see Fig.~\ref{fig:fisher}).
Furthermore at higher $z$ the non-linear scale $k_{\rm nl}$ moves out to higher $k$; 
a more detailed treatment would take $k_{\rm max} \propto k_{\rm nl}(z)$
\cite{BlaGla03}.
Translated into constraints on $w$, moving $k_{\rm max}=0.15$ Mpc$^{-1}$ 
to $0.075$ Mpc$^{-1}$ takes $\sigma(w)$ from $0.024$ to $0.051$ for
the galaxy survey and $0.04$ to $0.06$ for the cluster survey.

The redshift resolution required by the cluster survey is also less
stringent than for the galaxy survey.  Examination of Fig.~\ref{fig:fisher} shows
that most of the information on both $D_A$ and $H$ comes from modes with
$k_\parallel < 0.06$ Mpc$^{-1}$ ($\lambda_z = 0.03$ at $z=0.4$).  
For the cluster survey, excluding higher $k_\parallel$ degrades 
$D_A$ by $1.2$ and $H$ by $1.7$ for
the $z=0.3-0.5$ band and less for the higher redshift bands.
With an improvement in photometric redshift techniques and averaging over
cluster members,
costly spectroscopic followup could potentially be avoided.

The mass threshold and followup-redshift range of the cluster survey also 
affect potential constraints.  The rarity of high-$z$ clusters due to the
steepness of the mass function makes a low mass threshold crucial for 
the recovery of information at $z \sim 1$.  The requirement is less stringent
for measuring a constant $w$ since the lower redshift clusters suffice
in the determination.
With $M > 10^{14}$ $h^{-1} M_\odot$, $\sigma(w)$ improves by 1.5; with
$z < 0.7$ it degrades by 1.3. Increasing the sky coverage of the survey decreases the
errors as $A_s^{-1/2}$ to the ultimate limit of an improvement by a factor of $3$ for full-sky 
coverage in the cluster survey.

Removing the weak prior on $\beta$ $[\sigma(\ln \beta)=0.4]$ 
has a negligible effect on
galaxies constraints and causes a factor of $\sim 1.2$ degradation for the
cluster $D_A$ and $H$ constraints.
The weak prior on the overall bias $[\sigma(\ln b)=0.4]$ also has little
effect on the constraints.  An improvement in prior knowledge to the
percent level would substantially assist constraints due in part to the degeneracy
between the power spectrum normalization and the volume effects in the
shift (see discussion following Eqn.~[\ref{eqn:fisher}]).  
Likewise a further weakening
of the prior knowledge to account for bias evolution uncertainties by
considering the bias at each redshift slice as an independent parameter
degrades projections for $\sigma(w)$ by a factor of $1.6$ for the
galaxies and $1.9$ for the clusters as
shown in Fig.~\ref{fig:ellipse} (open ellipses).

\section{Discussion}
\label{sec:discussion}

The cosmic microwave background (CMB) provides a template
for power spectrum features in the linear regime, or equivalently a set of
absolutely calibrated standard rulers for cosmology.
We have provided simple tools to estimate cosmological information
contained therein. Angular diameter distances and the Hubble parameter
can be measured in a purely geometric way that involves 
only the well-understood physics of the CMB and linear perturbation theory.
A detection of the acoustic rings would assist in making the angular
diameter distance measurement robust, but is not strictly required if one assumes
continued improvement in the determination of the power spectrum template from
the CMB.  Conversely, current determinations of the power spectrum template
suffice if the acoustic rings are detected.

The tools we have introduced should
assist in planning future surveys to complement current cosmological
knowledge.  When actually analyzing data,
they of course must be replaced by more sophisticated tools that account
for cosmological evolution across the survey sub--volumes, the angular mask,
the radial selection, and curvature of the sky \cite{MatSza02}.  
The extent of the linear regime and non-linear corrections in the trans--linear
regime must also be addressed more carefully in simulations \cite{MeiWhiPea99}.

These estimates show that power spectrum rings at intermediate redshifts $z<1$ hold
great promise for cosmology.  
The angular distortion can be measured
with objects that are at least as abundant as clusters of galaxies.
At $z<0.4$ the ongoing SDSS LRG survey has the potential
to measure the Hubble constant, the best complement to the CMB, and potentially the dark
energy driven evolution of $D_A$ and $H$ separately
to obtain net constraints at the $\sigma(w)=0.03-0.05$
level, depending on the extent of the linear regime.  
At intermediate redshifts, clusters can probe the onset of dark energy 
domination with comparable precision $\sigma(w)=0.04-0.08$, ranging from
perfect constraints on bias evolution to no constraints.
Cluster distance and even crude Hubble parameter measurements 
do not require high resolution in redshift and are potentially
possible with photometric redshifts if the errors can be reduced to $\Delta z < 0.01$.  

By $z>1$
the expectations are that dark energy is sub--dominant and hence measurements become degenerate
with the information in the CMB.  A measurement 
would test that expectation and potentially reveal a more exotic form of
the dark energy \cite{Eis03}. 
Indeed measurement of the geometric shifting of power spectrum rings 
at any redshift
would complement luminosity distance ratios
from supernovae with a purely geometric test
and lend credence to any future detection of a dynamical dark energy component.

\smallskip
{\it Acknowledgments:} We thank
A. Berlind, D. Eisenstein, J. Frieman, D. Huterer, 
A.V. Kravtsov, J. Mohr, C. Pryke, E. Sheldon, D.N. Spergel and  I. Zehavi 
for useful conversations and the 
members of the DUET/DUO team for the initial motivation of this work. 
 WH is supported by NASA NAG5-10840 and
the DOE OJI program.

\end{document}